# First International Conference on Diffusion in Solids and Liquids

6-8 of July, 2005
Aveiro - Portugal

*Proceedings*

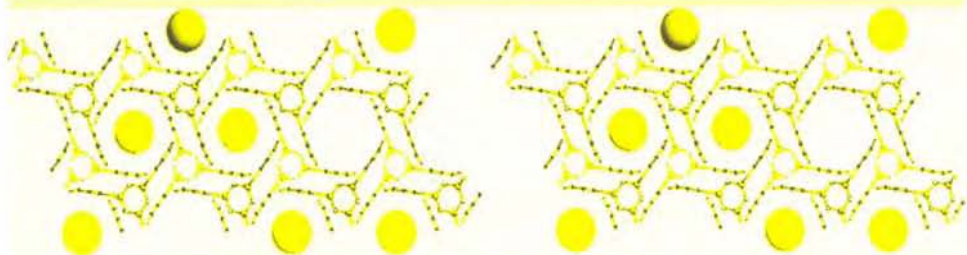

Editors:

**Andreas Öchsner**

**José Grácio**

**Frederic Barlat**

Proceedings of the
1st International Conference on Diffusion in Solids and Liquids
"DSL-2005"

Volume I

*Editors:*

Andreas Öchsner
José Grácio
Frédéric Barlat

Design – *Marcia Öchsner*

# MICROSCOPIC APPROACH TO THE EVALUATION OF DIFFUSION COEFFICIENTS FOR SUBSTITUTIONAL F.C.C. SOLID SOLUTIONS

Taras RADCHENKO[1], Valentyn TATARENKO[1,*], Sergiy BOKOCH[2], Mykola KULISH[2]

[1]Department of Solid State Theory, G. V. Kurdyumov Institute for Metal Physics, N.A.S. of Ukraine,
36 Academician Vernadsky Boulevard, 03680 Kyyiv-142, Ukraine
[2]Department of Physics of Functional Materials, Taras Shevchenko Kyyiv National University,
6 Academician Glushkov Prospekt, 03650 Kyyiv-22, Ukraine

[*]Corresponding author. Fax: +380 44 4242561; Email: tatar@imp.kiev.ua

**Abstract**

The microscopic theory of atomic diffusion kinetics is used for f.c.c. substitutional solid solutions. Within this approach, the short-range order relaxation is due to the atomic migration. Experimental data on the time dependence of radiation diffuse scattering are used for the determination of microscopic characteristics of atomic migration. The model takes into account the discrete and anisotropic character of atomic jumps in a long-range field of the concentration heterogeneities of interacting atoms. Such a consideration is applied for a close-packed Ni–Fe solid solution. Atomic-jumps' probabilities are estimated that allows to determine the diffusion coefficients and activation energies. Independent kinetic experimental data about a time evolution of long-range order are also used to calculate diffusivities in $L1_2$-Ni–Fe alloy.

## 1 Introduction

It is well known that existing experimental methods for studying diffusion processes are based on the measurements in solid solutions where concentration heterogeneities have macroscopic spatial dimensions greatly exceeding the lattice parameters. The information available from these experiments gives only macroscopic diffusivities for continuous diffusion equations. Besides, such experiments are characterized by low rates of diffusion relaxation of macroscopic concentration heterogeneities. To reduce the measurement time, the relatively high temperatures should be used. This is why the conventional data predominantly concern the high-temperature diffusion characteristics.

However, some diffusion processes occur within the atomic-scale ranges and can be studied by the radiation diffraction technique. They are: spinodal decomposition, long-range order kinetics below the stability limits of the disordered phase, homogenization of the sandwich-like deposit structures, short-range order kinetics, and so on [1]. The first three processes should be described by the discrete theory of single-site probability kinetics (the theory of discrete diffusion), while a kinetics of the short-range order should be described by the discrete theory of two-site probability kinetics.

In fact, short-range order is the unique natural occurring concentration heterogeneity, whose dimensions are commensurate with lattice parameters of a solid solution. Kinetics of short-range order is determined by the microscopic diffusion over intersite distances. Therefore, kinetic measurements of its relaxation provide us with detailed information on the discrete diffusion mechanism such as a possibility to determine the microscopic characteristics of atomic migrations, including probabilities and types of atomic jumps, and activation energy of diffusion.

The most convenient instrument for the investigation of short-range order kinetics, that is elementary diffusion events, is relaxation of radiation (X-ray, thermal neutrons) diffuse scattering intensities [1–3]. Besides, discrete diffusion measurements can be performed at room temperatures, because of short time of elementary diffusion events. This means that results can be utilized to determine the low-temperature diffusivities and activation energies.

The goal of a present paper is to estimate the microscopic characteristics of diffusion elementary events in Ni–Fe solid solution: 'potential'-fields' distribution due to the concentration heterogeneities caused by the atom at zero ('central') site and probabilities of atomic jumps per unite time at arbitrary sites of f.c.c. lattice. Diffusion parameters are determined using experimental data [3] on thermal neutrons' diffuse scattering within the framework of first-order kinetics model. Microscopic values enable to calculate the macroscopic characteristics: diffusion and self-diffusion coefficients. On the other hand, we use independent experimental data [4] on long-range order relaxation kinetics to estimate the diffusivities in this alloy.

## 2 Model of diffuse-scattering kinetics

Let us consider a two-particle correlation function $P_{\alpha\beta}(\mathbf{R} - \mathbf{R}', t)$—probability to find simultaneously (at a given time $t$) atoms of $\alpha$, $\beta$ components at the sites $\mathbf{R}$, $\mathbf{R}'$ ($\mathbf{R}$, $\mathbf{R}'$—radius-vectors of Bravais lattice sites). For a binary substitutional solid solution, time dependence of its Fourier transforms,

$$\widetilde{P}_{\alpha\beta}(\mathbf{k},t) = \sum_{\mathbf{R}-\mathbf{R}'} P_{\alpha\beta}(\mathbf{R}-\mathbf{R}',t) e^{-i\mathbf{k}\cdot(\mathbf{R}-\mathbf{R}')}, \quad (1)$$

can be approximately represented as follows [1]:

$$\Delta\widetilde{P}_{\alpha\beta}(\mathbf{k},t) = a_{11}^{\alpha\beta}(\mathbf{k}) e^{-2\lambda_1(\mathbf{k})t} + \\ + a_{22}^{\alpha\beta}(\mathbf{k}) e^{-2\lambda_2(\mathbf{k})t} + a_{12}^{\alpha\beta}(\mathbf{k}) e^{-[\lambda_1(\mathbf{k})+\lambda_2(\mathbf{k})]t}; \quad (2)$$

$\Delta\widetilde{P}_{\alpha\beta}(\mathbf{k},t) = \widetilde{P}_{\alpha\beta}(\mathbf{k},t) - \widetilde{P}_{\alpha\beta}(\mathbf{k},\infty)$; $\widetilde{P}_{\alpha\beta}(\mathbf{k},\infty)$ is equilibrium value; $a_{11}^{\alpha\beta}(\mathbf{k})$, $a_{22}^{\alpha\beta}(\mathbf{k})$, $a_{12}^{\alpha\beta}(\mathbf{k})$—pre-exponential coefficients; $\{2\lambda_1(\mathbf{k})\}^{-1}$, $\{2\lambda_2(\mathbf{k})\}^{-1}$, $\{\lambda_1(\mathbf{k}) + \lambda_2(\mathbf{k})\}^{-1}$ represent the possible relaxation times of three atoms'-configuration modes with a wave vector $\mathbf{k}$; $\lambda_1(\mathbf{k})$, $\lambda_2(\mathbf{k})$—eigenvalues of matrix

$$\widetilde{W}_{\alpha\beta}(\mathbf{k}) = \sum_{\gamma} [\widetilde{L}_{\alpha\gamma}(\mathbf{k})/k_B T] c_\alpha c_\gamma \widetilde{\Psi}_{\gamma\beta}(\mathbf{k}). \quad (3)$$

In the last equation, $\widetilde{L}_{\alpha\beta}(\mathbf{k})$ is Fourier transform of Önsager kinetic coefficient $L_{\alpha\beta}(\mathbf{R} - \mathbf{R}')$, which defines the exchange probability between a pair of $\alpha$ and $\beta$ atoms at lattice sites $\mathbf{R}$ and $\mathbf{R}'$ per unit time; $k_B$—Boltzmann constant; $T$—temperature; $c_\alpha$ is the atomic fraction of the $\alpha$-kind atoms; $\widetilde{\Psi}_{\alpha\beta}(\mathbf{k})$ is a Fourier transform of matrix

$$\Psi_{\alpha\beta}(\mathbf{R} - \mathbf{R}') \approx \left(\delta^2 F / [\delta P_\alpha(\mathbf{R}) \delta P_\beta(\mathbf{R}')]\right)_{\substack{P_\alpha(\mathbf{R})=c_\alpha \\ P_\beta(\mathbf{R}')=c_\beta}} \quad (4)$$

within the framework of the quasi-chemical approximation [1]. In Eq. (4), $F$—free energy of the equilibrium binary solid solution with one-particle probability $P_\alpha(\mathbf{R})$ to find the $\alpha$-kind atom at the site $\mathbf{R}$ (in disordered state $P_\alpha(\mathbf{R}) \equiv c_\alpha$).

Diffuse scattering of radiations is caused by the atoms'-configuration fluctuations, *i.e.* short-range order of atoms. Short-range order relaxation 'promotes' the diffuse-scattering relaxation. In a reciprocal space, distribution of diffuse scattering intensity, $I_{\text{diff}}(\mathbf{k},t)$, is connected with two-particle correlation function [1, 2]:

$$\Delta I_{\text{diff}}(\mathbf{k},t) = \sum_{\alpha,\beta} f_\alpha f_\beta \Delta\widetilde{P}_{\alpha\beta}(\mathbf{k},t); \quad (5)$$

$\Delta I_{\text{diff}}(\mathbf{k},t) = I_{\text{diff}}(\mathbf{k},t) - I_{\text{diff}}(\mathbf{k},\infty)$, $I_{\text{diff}}(\mathbf{k},\infty)$—diffuse scattering intensity in the equilibrium solid solution, wave vector $\mathbf{k}$ characterizes distance of the point of measurement from the nearest site of reciprocal lattice in $\mathbf{k}$-space of crystal, $f_\alpha$ and $f_\beta$—atomic-scattering factors of $\alpha$ and $\beta$ components.

Substituting Eq. (2) into (5) and assuming strict inequality $\lambda_1(\mathbf{k}) \ll \lambda_2(\mathbf{k})$, Eq. (5) results in

$$\Delta I_{\text{diff}}(\mathbf{k},t)/\Delta I_{\text{diff}}(\mathbf{k},0) \approx e^{-2\lambda_1(\mathbf{k})t}, \ \tau \approx 1/[2\lambda_1(\mathbf{k})]. \ (6)$$

This corresponds to the first-order kinetics model; $\tau$ is a relaxation time of intensity (in $\mathbf{k}$-space).

It is useful to express the Fourier component, $\lambda_1(\mathbf{k})$, in terms of function $\widetilde{\varphi}_1(\mathbf{k})$ [1] showing 'potential'-fields' action due to the concentration heterogeneities in solid solution:

$$\lambda_1(\mathbf{k}) \approx \lambda_1^0(\mathbf{k})[1 + \widetilde{\varphi}_1(\mathbf{k})]; \quad (7)$$

$\lambda_1^0(\mathbf{k})$ corresponds to the ideal solid solution. Function $\widetilde{\varphi}_1(\mathbf{k})$ is connected with Fourier transforms of interaction energies by means of $\widetilde{\Psi}_{\alpha\alpha}(\mathbf{k})$, $\widetilde{\Psi}_{\beta\beta}(\mathbf{k})$, $\widetilde{\Psi}_{\alpha\beta}(\mathbf{k})$ values [1] and is not vanishing in non-ideal solid solution only.

Thus, diffuse-scattering kinetics data enable to estimate the relaxation time $\tau$ and calculate $\lambda_1(\mathbf{k})$.

## 3 Determination of atomic jumps in f.c.c. lattice

Value of $\lambda_1(\mathbf{k})$ is also presented in equations of random-walk problem, which, in a case of long-wave approach, turn into diffusion equations. As shown in [1], using linear approximation within the framework of the first-order kinetics model, the time dependence of Fourier transform of one-particle probability $P_\alpha(\mathbf{R},t)$ (to find, at a time $t$, the atom $\alpha$ at a site $\mathbf{R}$ of the binary solution) can represented in a form of

$$\widetilde{P}_\alpha(\mathbf{k},t) \approx \widetilde{P}_\alpha(\mathbf{k},\infty) + a_1^\alpha(\mathbf{k}) e^{-\lambda_1(\mathbf{k})t}. \quad (8)$$

The last equation is valid for the real cases of substitutional diffusion, if the diffusion coefficients of components in binary solution are considerably different ($\lambda_1(\mathbf{k}) \ll \lambda_2(\mathbf{k})$). In such a case, one can suppose that 'fast' atoms (of $\beta$) form a quasi-equilibrium atmosphere around 'slow' atoms (of $\alpha$). Therefore, the time evolution of the $\alpha$-atoms should be only considered.

Krivoglaz [5] has considered the random walk of atoms. Solution of this problem has a form, which coincides with Eq. (8). In terms of [5], a value analogous with $\lambda_1(\mathbf{k})$ represents Fourier transform (with a sign 'minus') of atomic-jumps' probability for α component into a given site $\mathbf{R}$ in a unite time. Such a problem describes the process, which takes place in a 'gas' of non-interacting atoms and is connected with phenomenon of self-diffusion. So, if, for any $\mathbf{k}$ in Brillouin zone, $\widetilde{\varphi}_1(\mathbf{k}) \equiv 0$, then $-\lambda_1^0(\mathbf{k})$ is a Fourier transform of α-atom-jumps' probability per unite time over sites $\{\mathbf{R}\}$ in ideal solid solution.

As follows from Eq. (7), the same equation describes random walks of atoms in non-ideal and ideal solid solutions, if formally assume that $-\lambda_1(\mathbf{k})$ is Fourier transform of jumps' probability for α-atoms over sites $\{\mathbf{R}\}$ per unite time.

If $\{-\lambda_1(\mathbf{k})\}$ are known, one can obtain their Fourier original, $-\Lambda_\alpha(\mathbf{R})$—jumps' probability of α-atoms per unit time into the given site $\mathbf{R}$ from the all surroundings sites $\{\mathbf{R}'\}$ in a field of interaction 'potential' $\psi_\alpha(\mathbf{R}')$. We assume that 'potential' $\psi_\alpha(\mathbf{R}')$ at the site $\mathbf{R}'$ is generated, if microscopic concentration heterogeneities take place in a solid solution, for instance, because of location of α-atom at the 'zero' site. Apparently, $\Lambda_\alpha(\mathbf{R})$ for any $\mathbf{R}$ (including a 'zero' site) of the lattice is proportional to values of all $\{\psi_\alpha(\mathbf{R}')\}$. Using inverse Fourier transformation into the $\mathbf{R}$-space [6], $\Lambda_\alpha(\mathbf{R})$ can be written as:

$$\Lambda_\alpha(\mathbf{R}) \cong \sum_{\mathbf{R}'} \Lambda_\alpha^0(\mathbf{R}-\mathbf{R}')\psi_\alpha(\mathbf{R}')[c_\alpha(1-c_\alpha)/k_B T], \quad (9)$$

where $-\Lambda_\alpha^0(\mathbf{R}-\mathbf{R}')$—jump probability of α-atom into the site $\mathbf{R}$ from any site $\mathbf{R}'$ per unit time. At a 'zero' site of the ideal solid solution, 'potential' is $\psi_\alpha(\mathbf{0}) \approx \psi_0 \equiv k_B T/c_\alpha(1-c_\alpha)$. Value of $\Lambda_\alpha(\mathbf{R})$ is dependent on the location of sites in a crystal of a given syngony, *i.e.* on the set of possible differences $\{\mathbf{R}-\mathbf{R}'\}$ for each $\mathbf{R}$. We will also assume that region of 'potential'-field influence is bounded. This means that values of $\psi_\alpha(\mathbf{R})$ are non-zero for some co-ordination shells only.

In case of the vacancy mechanism of diffusion, we can take into account atomic jumps only over the nearest distances between the sites. When it is necessary to test the possibility of not only vacancy diffusion mechanism, but also more complex diffusion processes, then we have to consider several sets of values $\Lambda_\alpha^0(\mathbf{R}-\mathbf{R}'_{\text{I}})$, $\Lambda_\alpha^0(\mathbf{R}-\mathbf{R}'_{\text{II}})$, *etc.* Indexes I, II, *etc.* relate to the jumps to the site $\mathbf{R}$ from the nearest-neighbour sites $\{\mathbf{R}'_{\text{I}}\}$, next-nearest-neighbour sites $\{\mathbf{R}'_{\text{II}}\}$, *etc.* (Fig. 1).

Analogous models have been considered in [6–8]. In Ref. [6], authors assumed atomic jumps within the first co-ordination shell only; in Ref. [7], scheme of studying microdiffusion has been proposed only, without estimation of diffusivities.

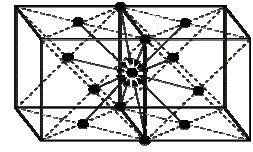

Figure 1: Atomic jumps into one of sites from nearest sites and next-nearest sites in f.c.c. lattice.

Let us assume that 'potential' field extends on the six co-ordination shells around the 'central' ('zero') site, and jump of α-atom is possible within the two co-ordination shells:

$$\Lambda_\alpha^0(R_{\text{I}}) = \Lambda_{\alpha\text{I}}^0 \neq 0; \quad \Lambda_\alpha^0(R_{\text{II}}) = \Lambda_{\alpha\text{II}}^0 \neq 0;$$
$$\psi_\alpha(R_{\text{I}}) = \psi_{\alpha\text{I}} \neq 0; \quad \psi_\alpha(R_{\text{II}}) = \psi_{\alpha\text{II}} \neq 0;$$
$$\psi_\alpha(R_{\text{III}}) = \psi_{\alpha\text{III}} \neq 0; \quad \psi_\alpha(R_{\text{IV}}) = \psi_{\alpha\text{IV}} \neq 0;$$
$$\psi_\alpha(R_{\text{V}}) = \psi_{\alpha\text{V}} \neq 0; \quad \psi_\alpha(R_{\text{VI}}) = \psi_{\alpha\text{VI}} \neq 0;$$

$R_{\text{I}}$, $R_{\text{II}}$, *etc.* are radii of the first, second, *etc.* co-ordination shells, respectively. If α-atom is located at the 'zero' site, other probabilities ($\Lambda_{\alpha\text{III}}^0$, $\Lambda_{\alpha\text{IV}}^0$, *etc.*) and 'potential' functions ($\psi_{\alpha\text{VII}}$, $\psi_{\alpha\text{VIII}}$, *etc.*) equal zero, then we can write the following equations with $\Lambda_\alpha(\mathbf{R}_n(lmn))$ ($lmn$—co-ordinates of sites in a usual system with basis vectors along [100], [010], [001] directions) for 8 nearest (with respect to 'zero' site) co-ordination shells only and for $\mathbf{R} = \mathbf{R}_0 = \mathbf{0}$ as well;

$$\Lambda_\alpha(\mathbf{R}_0(000)) \cong 12\Lambda_{\alpha\text{I}}^0[\psi_{\alpha\text{I}}/\psi_{\alpha 0}] +$$
$$+ 6\Lambda_{\alpha\text{II}}^0[\psi_{\alpha\text{II}}/\psi_{\alpha 0}] + \Lambda_{\alpha 0}^0,$$

$$\Lambda_\alpha(\mathbf{R}_{\text{I}}(\tfrac{1}{2}\tfrac{1}{2}0)) \cong \Lambda_{\alpha\text{I}}^0 + 4\Lambda_{\alpha\text{I}}^0[\psi_{\alpha\text{I}}/\psi_{\alpha 0}] +$$
$$+ 2\Lambda_{\alpha\text{I}}^0[\psi_{\alpha\text{II}}/\psi_{\alpha 0}] + \Lambda_{\alpha\text{I}}^0[\psi_{\alpha\text{III}}/\psi_{\alpha 0}] + 2\Lambda_{\alpha\text{I}}^0[\psi_{\alpha\text{II}}/\psi_{\alpha 0}] +$$
$$+ 4\Lambda_{\alpha\text{I}}^0[\psi_{\alpha\text{IV}}/\psi_{\alpha 0}] + \Lambda_{\alpha 0}^0[\psi_{\alpha\text{I}}/\psi_{\alpha 0}],$$

$$\Lambda_\alpha(\mathbf{R}_{\text{II}}(100)) \cong 4\Lambda_{\alpha\text{I}}^0[\psi_{\alpha\text{I}}/\psi_{\alpha 0}] + \Lambda_{\alpha\text{II}}^0 +$$
$$+ 4\Lambda_{\alpha\text{II}}^0[\psi_{\alpha\text{III}}/\psi_{\alpha 0}] + 4\Lambda_{\alpha\text{I}}^0[\psi_{\alpha\text{IV}}/\psi_{\alpha 0}] +$$
$$+ \Lambda_{\alpha 0}^0[\psi_{\alpha\text{II}}/\psi_{\alpha 0}] + \Lambda_{\alpha\text{II}}^0[\psi_{\text{VI}}/\psi_{\alpha 0}],$$

$$\Lambda_\alpha(\mathbf{R}_{\text{III}}(1\tfrac{1}{2}\tfrac{1}{2})) \cong \Lambda_{\alpha\text{I}}^0[\psi_{\alpha\text{I}}/\psi_{\alpha 0}] + 2\Lambda_{\alpha\text{I}}^0[\psi_{\alpha\text{III}}/\psi_{\alpha 0}] +$$
$$+ 2\Lambda_{\alpha\text{I}}^0[\psi_{\alpha\text{I}}/\psi_{\alpha 0}] + \Lambda_{\alpha\text{II}}^0[\psi_{\alpha\text{I}}/\psi_{\alpha 0}] + 2\Lambda_{\alpha\text{I}}^0[\psi_{\alpha\text{IV}}/\psi_{\alpha 0}] +$$
$$+ 2\Lambda_{\alpha\text{II}}^0[\psi_{\alpha\text{IV}}/\psi_{\alpha 0}] + \Lambda_{\alpha\text{I}}^0[\psi_{\alpha\text{V}}/\psi_{\alpha 0}] + \Lambda_{\alpha 0}^0[\psi_{\alpha\text{IV}}/\psi_{\alpha 0}],$$

$$\Lambda_\alpha(\mathbf{R}_{\text{IV}}(110)) \cong \Lambda_{\alpha\text{I}}^0[\psi_{\alpha\text{I}}/\psi_{\alpha 0}] + 2\Lambda_{\alpha\text{II}}^0[\psi_{\alpha\text{II}}/\psi_{\alpha 0}] +$$
$$+ 4\Lambda_{\alpha\text{I}}^0[\psi_{\alpha\text{IV}}/\psi_{\alpha 0}] + 2\Lambda_{\alpha\text{I}}^0[\psi_{\alpha\text{V}}/\psi_{\alpha 0}] + \Lambda_{\alpha 0}^0[\psi_{\alpha\text{III}}/\psi_{\alpha 0}],$$

$$\Lambda_\alpha(\mathbf{R}_{\text{V}}(1\tfrac{1}{2}\tfrac{1}{2}0)) \cong \Lambda_{\alpha\text{II}}^0[\psi_{\alpha\text{II}}/\psi_{\alpha 0}] + \Lambda_{\alpha\text{I}}^0[\psi_{\alpha\text{III}}/\psi_{\alpha 0}] +$$
$$+ \Lambda_{\alpha\text{II}}^0[\psi_{\alpha\text{I}}/\psi_{\alpha 0}] + \Lambda_{\alpha\text{I}}^0[\psi_{\alpha\text{IV}}/\psi_{\alpha 0}] + \Lambda_{\alpha\text{I}}^0[\psi_{\alpha\text{VI}}/\psi_{\alpha 0}],$$

$$\Lambda_\alpha(\mathbf{R}_{\text{VI}}(111)) \cong 3\Lambda_{\alpha\text{II}}^0[\psi_{\alpha\text{III}}/\psi_{\alpha 0}] +$$

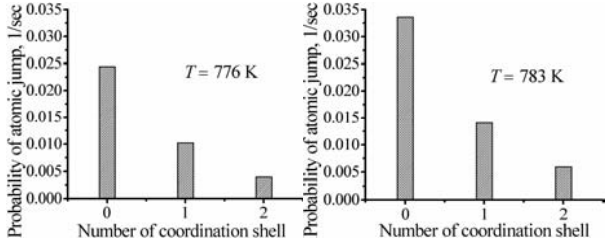

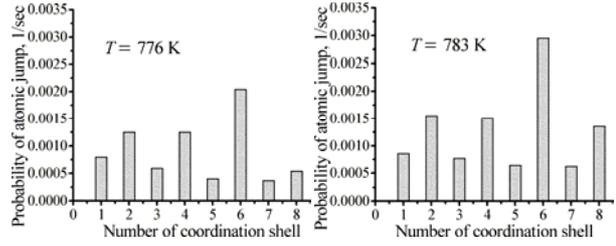

Figure 2: Jumps' probabilities for α-atoms within the two co-ordination shells (numbered by n = 0, 1, 2) per unite time, $-\Lambda_\alpha^0(R_n)$, in 'ideal' f.c.c.-Ni$_{0.765}$Fe$_{0.235}$-type solution.

Figure 4: Jumps' probabilities of α-atoms into the site $\mathbf{R}_n$ per unite time, $-\Lambda_\alpha(R_n)$ (n—number of co-ordination shell), in non-ideal disordered f.c.c.-$^{62}$Ni$_{0.765}$Fe$_{0.235}$.

$$+ 3\Lambda_{\alpha I}^0[\psi_{\alpha IV}/\psi_{\alpha 0}] + \Lambda_{\alpha 0}^0[\psi_{\alpha V}/\psi_{\alpha 0}],$$

$$\Lambda_\alpha(\mathbf{R}_{VII}(1½\,1\,½)) \cong \Lambda_{\alpha I}^0[\psi_{\alpha III}/\psi_{\alpha 0}] + \Lambda_{\alpha I}^0[\psi_{\alpha IV}/\psi_{\alpha 0}] +$$
$$+ \Lambda_{\alpha II}^0[\psi_{\alpha IV}/\psi_{\alpha 0}] + \Lambda_{\alpha I}^0[\psi_{\alpha V}/\psi_{\alpha 0}],$$

$$\Lambda_\alpha(\mathbf{R}_{VIII}(200)) \cong \Lambda_{\alpha I}^0[\psi_{\alpha VI}/\psi_{\alpha 0}] + \Lambda_{\alpha II}^0[\psi_{\alpha II}/\psi_{\alpha 0}];$$

here $-\Lambda_{\alpha 0}^0$—probability of α-atom to stay *in situ* (at a given site) per unite time, $\psi_{\alpha 0}$—'potential' function at the 'zero' site. Values of $\Lambda_\alpha(\mathbf{R}_n(lmn))$ are obtained for disordered f.c.c.-$^{62}$Ni$_{0.765}$Fe$_{0.235}$ solution from inverse Fourier transformation using experimental data [3] by the estimated values of $\lambda_1(\mathbf{k})$ on a basis of kinetic model (6). Fourier original of probability to jump into the site $\mathbf{R}$ of a cubic lattice is as follows:

$$\Lambda_\alpha(\mathbf{R}) = K(lmn) \times$$
$$\times \sum_{h_1 h_2 h_3} \lambda_1(h_1 h_2 h_3) \cos(2\pi h_1 l) \cos(2\pi h_2 m) \cos(2\pi h_3 n),$$

where $K$—geometrical coefficient, which is dependent on the co-ordinates of vector $\mathbf{R}(lmn)$.

Jumps' probabilities of 'slow' α-atoms (*i.e.* conditionally 'slow' Fe atoms within the 'coat' of 'fast' Ni atoms [9]) within the first two co-ordination shells in 'ideal' Ni$_{0.765}$Fe$_{0.235}$-type alloy for temperatures 776 K and 783 K are presented in Fig. 2. The first column in both figures means a probability (per unite time) of 'slow' atom to stay *in situ* at a given site. Magnitude of this probability ($\approx 0.024$ sec$^{-1}$ and 0.033 sec$^{-1}$ for 776 K and 783 K, respectively), just as two another probabilities for the first and second co-ordination shells, is, for example, smaller than for f.c.c.-Ni–Mo ($\approx 0.6$ sec$^{-1}$ [8]).

Caused by microscopic heterogeneity, 'potential' function of $R_n$ (Fig. 3) determines the atomic jumps (Fig. 4) during short-range order relaxation. It has oscillating character, and its absolute value decreases as a whole with increasing distance from the concentration heterogeneity at the 'zero' site (Fig. 3).

## 4 Calculation of diffusivities

Atomic-jumps' probabilities enable to calculate macroscopic diffusion characteristics, *i.e.* diffusion and self-diffusion coefficients of 'slow' α-atoms.

Equation (7) is analogue to the Darken formula:

$$D_\alpha = D_\alpha^*[1 + (d \ln \gamma_\alpha)/(d \ln c_\alpha)]; \qquad (10)$$

here $D_\alpha, D_\alpha^*, \gamma_\alpha$—diffusion, self-diffusion and activity coefficients, respectively, for α-atoms. So realising long-wave limit transition from discrete to continual process and comparing Eq. (7) with Eq. (10), we obtain the following relationship between the diffusion micro- and macroparameters:

$$\lambda_1(\mathbf{k}) \approx D_\alpha |\mathbf{k}|^2, \lambda_1^0(\mathbf{k}) \approx D_\alpha^* |\mathbf{k}|^2, \widetilde{\varphi}_\alpha(\mathbf{k}) \cong \frac{d \ln \gamma_\alpha}{d \ln c_\alpha} \quad (11)$$

(for $\mathbf{k} \to \mathbf{0}$) neglecting Kirkendall effect, at least. By definition, value $\lambda_1(\mathbf{k})$ is connected with jumps' probability $-\Lambda_\alpha(\mathbf{R}_n)$ of α-atoms per unit time:

$$\lambda_1(\mathbf{k}) = -\sum_{n=1}^\infty \sum_{\mathbf{R}_n} \Lambda_\alpha(\mathbf{R}_n)[1 - \exp(-i\mathbf{k} \cdot \mathbf{R}_n)]. \quad (12)$$

For $\mathbf{k} \to \mathbf{0}$, expanding $\exp(-i\mathbf{k} \cdot \mathbf{R}_n)$ in a series and retaining first nonzero terms ($\lambda_1(\mathbf{0}) = 0$), we obtain [1]:

$$\lambda_1(\mathbf{k}) \approx -(1/6) \sum_{n=1}^\infty \Lambda_\alpha(R_n) R_n^2 Z_n |\mathbf{k}|^2, \quad (13)$$

$Z_n$—co-ordination number for a n-th co-ordination shell. Comparison of Eq. (13) with (11) yields formula for diffusivity in non-ideal cubic solution:

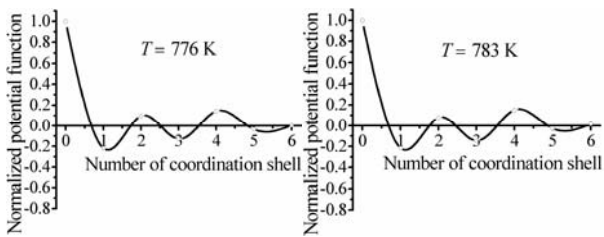

Figure 3: Normalized values of 'potential' functions, $\psi_\alpha(R_n)/\psi_{\alpha 0}$ (produced by the α-atom at the 'zero' site) at different co-ordination shells (n = I, …, VI) in f.c.c.-$^{62}$Ni$_{0.765}$Fe$_{0.235}$ for two temperatures.

Table 1: Vacancy-controlled diffusion ($D_{Fe}$), self-diffusion ($D^*_{Fe}$), and interdiffusion ($D$) coefficients for $^{62}$Ni$_{0.765}$Fe$_{0.235}$.

| $T$, K | $D_{Fe}$, cm$^2$/sec | $D^*_{Fe}$, cm$^2$/sec | $D$, cm$^2$/sec [3] |
|---|---|---|---|
| 776 | $4.49 \cdot 10^{-17}$ | $1.81 \cdot 10^{-17}$ | $2.49 \cdot 10^{-18}$ |
| 783 | $6.90 \cdot 10^{-17}$ | $2.55 \cdot 10^{-17}$ | $3.56 \cdot 10^{-18}$ |

$$D_\alpha \approx -(1/6)\sum_{n=1}^{\infty} \Lambda_\alpha(R_n) R_n^2 Z_n. \quad (14)$$

Similarly, for ideal solid solution,

$$D^*_\alpha \approx -(1/6)\sum_{n=1}^{\infty} \Lambda^0_\alpha(R_n) R_n^2 Z_n. \quad (15)$$

In a given model, we assume that atomic jumps into the sites within the same co-ordination shell (with respect to the 'zero' site) are equiprobable.

Diffusion and self-diffusion coefficients of Fe atoms in f.c.c.-$^{62}$Ni$_{0.765}$Fe$_{0.235}$ solution calculated from Eqs. (14), (15) are listed in Table 1. In the last column, interdiffusion coefficients extrapolated from the high-temperature measurements [3] are presented too. Total activation energies of vacancy-mediated diffusion and self-diffusion of Fe atoms are also calculated: 2.13 and 3.47 eV, respectively.

Now let us consider the case of exchange mechanism governing the diffusion in Ni$_3$Fe alloy (which is similar by the composition to Ni$_{0.765}$Fe$_{0.235}$ solid solution) at temperatures below the temperature of order–disorder phase transformation. Atomic distribution function in this ordering alloy can be represented as a superposition of the static concentration waves (SCW) [1]. Using SCW approach along with self-consistent-field approximation, one can consider the differential kinetic equation for relaxation of long-range order parameter $\eta$ [1] of $L1_2$-type Ni$_3$Fe:

$$\frac{d\eta}{dt} = \frac{3\tilde{L}_{\alpha\alpha}(\mathbf{k}_X)}{16} \left[ \frac{\eta \tilde{w}(\mathbf{k}_X)}{k_B T} + \ln \frac{(1+3\eta)(3+\eta)}{3(1-\eta)^2} \right]. \quad (16)$$

In the last equation, $\mathbf{k}_X = (100)$—wave vector (in a first Brillouin zone of f.c.c. lattice), which generates the $L1_2$-type superstructure, $\tilde{L}_{\alpha\alpha}(\mathbf{k}_X)$—Fourier transform of Önsager kinetic coefficients, $\tilde{w}(\mathbf{k}_X)$—Fourier component of pairwise-interaction 'mixing' energy between Ni and Fe atoms; magnitudes of $\tilde{w}(\mathbf{k}_X)$ for Ni$_3$Fe alloy with different temperatures were estimated in [7]. An implicit solution $t = t(\eta)$ of Eq. (16) is given by the following equation:

$$\int_{\eta_0}^{\eta} d\eta \left[ \frac{\eta \tilde{w}(\mathbf{k}_X)}{k_B T} + \ln \frac{(1+3\eta)(3+\eta)}{3(1-\eta)^2} \right]^{-1} = \frac{3\tilde{L}_{\alpha\alpha}(\mathbf{k}_X) t}{16}, \quad (17)$$

where $\eta_0$—initial magnitude of the long-range order parameter (at $t = 0$). Assuming atomic jumps between nearest-neighbour sites only and using

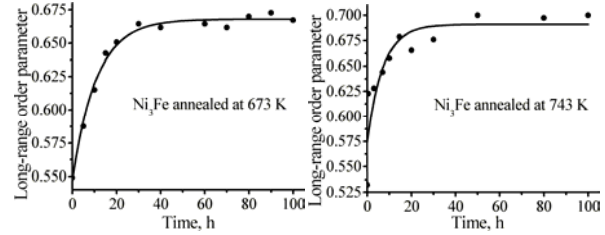

Figure 5: Time dependence of long-range order parameter in $L1_2$-type Ni$_3$Fe alloy; ●—experimental data [4].

condition that the total number of atoms in a system is conserved, for f.c.c. lattice, we can write

$$\tilde{L}_{\alpha\alpha}(\mathbf{k}) \approx -4L_{\alpha\alpha}(R_I)[3 - \cos\pi h_1 \cdot \cos\pi h_2 - \cos\pi h_2 \cdot \cos\pi h_3 - \cos\pi h_3 \cdot \cos\pi h_1], \quad (18)$$

where $L_{\alpha\alpha}(R_I)$ is proportional to the jump probability of a pair of atoms at nearest-neighbour sites $\mathbf{R}$ and $\mathbf{R}'$ per unite time ($R_I = |\mathbf{R} - \mathbf{R}'|$). Using experimental data [4] (Fig. 5) and Eqs. (17), (18), we obtain roughly averaged values: $\langle L_{\alpha\alpha}(R_I) \rangle \approx 4 \cdot 10^{-8}$ sec$^{-1}$ for $T = 673$ K and $\langle L_{\alpha\alpha}(R_I) \rangle \approx 7 \cdot 10^{-8}$ sec$^{-1}$ for $T = 743$ K. Substituting $L_{\alpha\alpha}(R_I)$ instead of $-\Lambda^0_\alpha(R_n)$ in equation similar to Eq. (15), exchange diffusive mobilities of 'slow' Fe atoms in $L1_2$-type Ni$_3$Fe alloy may be coarsely estimated by averaging data of Table 2 over annealing times: $\langle D^0_{Fe} \rangle \approx 1.03 \cdot 10^{-22}$ cm$^2$/sec for $T = 673$ K and $\langle D^0_{Fe} \rangle \approx 1.78 \cdot 10^{-22}$ cm$^2$/sec for $T = 743$ K. Accordingly estimated diffusion migration energy of Fe atoms is 0.34 eV.

## 5 Discussion of results and conclusions

Obtained microparameters of diffusion, $-\Lambda^0_\alpha(R_n)$ and $-\Lambda_\alpha(R_n)$, are estimated on the basis of available experimental data about diffuse-scattering kinetics, which is caused by the short-range order relaxation in the solid solution. Thus, obtained 'microdiffusivities' determine the short-range order changes and give information about their kinetics.

For 'ideal' f.c.c.-Ni$_{0.765}$Fe$_{0.235}$-type solid solution, probability of atomic jumps into a given site $\mathbf{R}$ of the nearest-neighbour shell around a 'zero'-site heterogeneity is less than $-\Lambda_\alpha(\mathbf{0})/2$, and of the next-nearest-neighbour sites—considerably less than $-\Lambda_\alpha(\mathbf{0})/2$ (Fig. 2). This stands for the predominance of atomic jumps within the first co-ordination shell that is mainly governed by the vacancy mechanism of diffusion within the commonly accepted interpretation.

Table 2: Exchange diffusive mobilities in $L1_2$-type Ni$_3$Fe alloy for different annealing times and temperatures.

| | $T = 673$ K | $T = 743$ K |
|---|---|---|
| $t$, h | $D^0_{Fe}$, cm$^2$/sec | $D^0_{Fe}$, cm$^2$/sec |
| 100 | $8.99 \cdot 10^{-23}$ | $1.26 \cdot 10^{-22}$ |
| 90 | $1.04 \cdot 10^{-22}$ | $1.55 \cdot 10^{-22}$ |
| 80 | $1.15 \cdot 10^{-22}$ | $2.52 \cdot 10^{-22}$ |

Dependence of normalized 'potential' function on a radius of co-ordination shell, $R_n$, is non-monotone (Fig. 3). For some $R_n$, function value is positive, for another one—negative. This determines thermodynamic 'disadvantage' or 'advantage' of 'slow' α-atom to stay in corresponding sites $\{R_n\}$.

Probability of an α-atom jump into the site **R**, $-\Lambda_\alpha(\mathbf{R})$, in a field caused by the α-atom excess at the 'zero' site is determined by the value of this field at a site **R**. That is why the jumps' probability of α-atoms into the site **R** is non-monotone (Fig. 4): it is higher at sites, where arrangement of α-atom is more energy-wise preferable.

If we put Fe-atom at a cube corner of an f.c.c.-lattice conditional unit cell, then, in a presence of the short-range order of $L1_2$(Ni$_3$Fe)-type, Fe-atoms will try to occupy predominantly the cube corners (*viz.* sites within the II-nd, IV-th, VI-th, VIII-th co-ordination shells around a 'zero' site, which coincides with one of the cube corners), and Ni-atoms will be localized at the face centres of cubic unit cells. So, in our case, a short-range order of $L1_2$-type is characterized by the arrangement of Ni atoms at the face centres that is more energy-wise preferable for Ni. Evidently, that is why jumps' probabilities of α-atoms (Fe) into the sites $\{R_{II}\}$, $\{R_{IV}\}$, $\{R_{VI}\}$, $\{R_{VIII}\}$ (Fig. 4) are higher than jumps' probabilities into the sites $\{R_I\}$, $\{R_{III}\}$, $\{R_V\}$, and $\{R_{VII}\}$. These jumps into the sites of the II-nd, IV-th, VI-th, VIII-th co-ordination shells with respect to the cube-corner 'zero' site have to be realised mainly as nearest-distance jumps from the sites, where arrangement of α-atoms (Fe) is less energy-wise 'advantageous', that is from the sites of the I-st, III-rd, V-th, VII-th co-ordination shells with respect to the cube-corner 'zero' site. Thus, n-dependent probabilities presented in Fig. 4 do not contrary to the vacancy mechanism of diffusion in alloy at issue.

It is quite clear that, if temperature increases, all probabilities increase as well (Figs. 2, 4).

Total activation energies of diffusion and self-diffusion of 'slow' Fe atoms in disordered alloy $^{62}$Ni$_{0.765}$Fe$_{0.235}$ are 2.13 eV and 3.47 eV, respectively. Exchange-diffusion migration energy of Fe atoms in long-range ordered Ni$_3$Fe proves to be 0.34 eV. The first two values are higher than the third one because the latter does not involve the energy of vacancy formation; migration energy in Ni$_3$Fe is evaluated within the alloy model without vacancies. Vacancy formation energy in a former model is 84–90%.

Diffusive mobilities listed in Table 2 are lower than diffusivities in Table 1 because of some reasons. Firstly, because of temperature- and concentration-dependent statistical-thermodynamic factor (of $\{c_\alpha c_\gamma \tilde{\Psi}_{\gamma\beta}(\mathbf{k})/(k_B T)\}$-type; see Eqs. (3), (9)). Secondly, below the order–disorder transformation temperature, the diffusion mechanism in long-range ordering alloys may be modified, and this will affect the value of $D$ in the direction of observed variation; in fact, the probability of exchange ('ring') mechanism of diffusion is small; it is proved by the magnitude of Önsager kinetic coefficient. Thirdly, because of temperature is decreased.

A given paper shows the possibility evaluating 'macrodiffusivities' by means of 'microdiffusivities' from the independent data on short- or long-range order kinetics. The presented scheme can be used for multicomponent systems based on f.c.c. lattice as well.

## Acknowledgements

One of authors (T.R.) is grateful for the NATO Reintegration Grant support under reference RIG 981326.

## References

[1] Khachaturyan, A.G.: *Theory of Structural Transformations In Solids.* John Wiley & Sons, New York, 1983.
[2] Krivoglaz, M.A.: *Diffuse Scattering of X-Rays and Neutrons—Fluctuations in Solids.* Springer, Berlin, 1996.
[3] Bley, F., Amilius, Z., Lefebvre, S *Acta metall.*, 36, pp. 1643–1652, 1988.
[4] Goman'kov, V.I., Puzey, I.M., Rukosuev, M.N. *Metallofizika,* Vol. 20. Naukova Dumka, Kiev, 1968.
[5] Krivoglaz, M.A. *Zh. Ehksp. Teor. Fiz.*, 40, pp. 1812–1824, 1961.
[6] Naumova, M.M., Semenovskaya, S.V., Umanskiy, Ya.S. *Fiz. Tverd. Tela,* 12, pp. 975–982, 1970.
[7] Tatarenko, V.A., Radchenko, T.M. *Metallofiz. Noveishie Tekhnol.*, 24, pp. 1335–1350, 2002.
[8] Bokoch, S.M., Kulish, M.P., Tatarenko, V.A., Radchenko, T.M. *Metallofiz. Noveishie Tekhnol.*, 26, pp. 541–558, 2004.
[9] Caplain, A., Chambron, W. *Acta metall.*, 25, p. 1001, 1977.